\documentclass[doublecol]{epl2}
\title{Equal partners do better in defensive alliances}
\shorttitle{Equal partners do better in defensive alliances}

\author{Marcell Blahota, Istv{\'a}n Blahota, and Attila Szolnoki}
\shortauthor{Blahota et al.}
\institute{Institute of Technical Physics and Materials Science, Centre for Energy Research, Hungarian Academy of Sciences, P.O. Box 49, H-1525 Budapest, Hungary}

\abstract{Cyclic dominance offers not just a way to maintain biodiversity, but also serves as a sort of defensive alliance against an external invader. Interestingly, a new level of competition can be observed when two cyclic loops are present. Here the inner invasion speed plays a decisive role on the evolutionary outcome, because faster invasion rate provides an evolutionary advantage to an alliance. In this Letter we demonstrate that the heterogeneity of inner invasion rates makes an alliance vulnerable against a loop where group members are equal. Quite surprisingly, a loop where invasion rates are uniform can still dominate an alliance formed by heteregeneous rates even if the average speed of invasion is significantly higher in the latter group. At a specific range of parameter space, when intergroup invasion or the average inner invasion is moderate, the heterogeneous alliance with higher internal invasion speed may prevail, or the system terminates onto a novel 4- or 5-species solution.}

\begin{document}

\maketitle
To explain diversity based on the Darwinian selection principle is a long standing challenge for biology, ecology and social sciences \cite{kerr_n02, szolnoki_jrsif14, baker_jtb20, frey_pa10, mathiesen_prl11, nagatani_srep18, dobramysl_jpa18, park_c18, avelino_pre19b}. Cyclic dominance between competitors is a well-known mechanism to maintain it, because in this case there is not an unambiguous victor and every competitor becomes predator or prey in different interactions. The simplest form of such kind of non-transitive interaction is the celebrated rock-scissors-paper game where paper invades rock, rock outperforms scissors, and scissors in turn dominate over paper. Interestingly, the essence of this cyclic dominance can be identified in several living systems ranging from plants, animals to bacterial communities \cite{watt_je47,lankau_s07, jackson_pnas75, sinervo_n96}. Furthermore, similar non-transitive relations may emerge in evolutionary game theory models where at least three or more strategies compete. Examples are social dilemmas with volunteering \cite{hauert_s02, semmann_n03}, punishment, reward \cite{helbing_ploscb10,sigmund_pnas01,szolnoki_epl10}, or the presence of destructive agents \cite{requejo_pre12}.

One may think that the vicinity of a predator or dominant strategy is always harmful to a prey or inferior strategy, but it is not necessarily true in a defensive alliance which is based on the above mentioned cyclic dominance \cite{szabo_pr07}. For example if $A$, $B$, and $C$ are members of a closed loop, and a fourth external $E$ beats $B$, while $A$ is a predator of $E$ and $B$ as well, then $A$ will beat both $B$ and $E$. Hence the cyclic loop can crowd out the intruder species.

More interestingly, not only species, but also alliances may compete \cite{roman_jtb16}. In the simplest version of such higher level competition we have six species arranged by a Lotka-Volterra type global cycle where two groups of three species form a closed loop separately and these two alliances struggle for space \cite{szabo_pre01b}. In a previous work it was shown that the speed of inner invasion within a cycle is an essential factor for the viability of an alliance. More precisely, the loop where the inner invasion is faster can dominate the other alliance where this inner rotation is slower \cite{perc_pre07b}. Interestingly, this mechanism can be detected in more realistic evolutionary game models, where the relation of competing strategies is less straightforward \cite{szolnoki_epl15}.

Earlier it was assumed that the invasion rate is uniform within a certain alliance, but this assumption is not really fulfilled in realistic systems. On the contrary, it is more common that the strengths of invasions within the circle are unequal \cite{tainaka_epl91}. This observation raises the question what if the invasion rates are not uniform? Is there any consequence on the vitality of a defective alliance?

\begin{figure}
\begin{center}
\includegraphics[width=6.0cm]{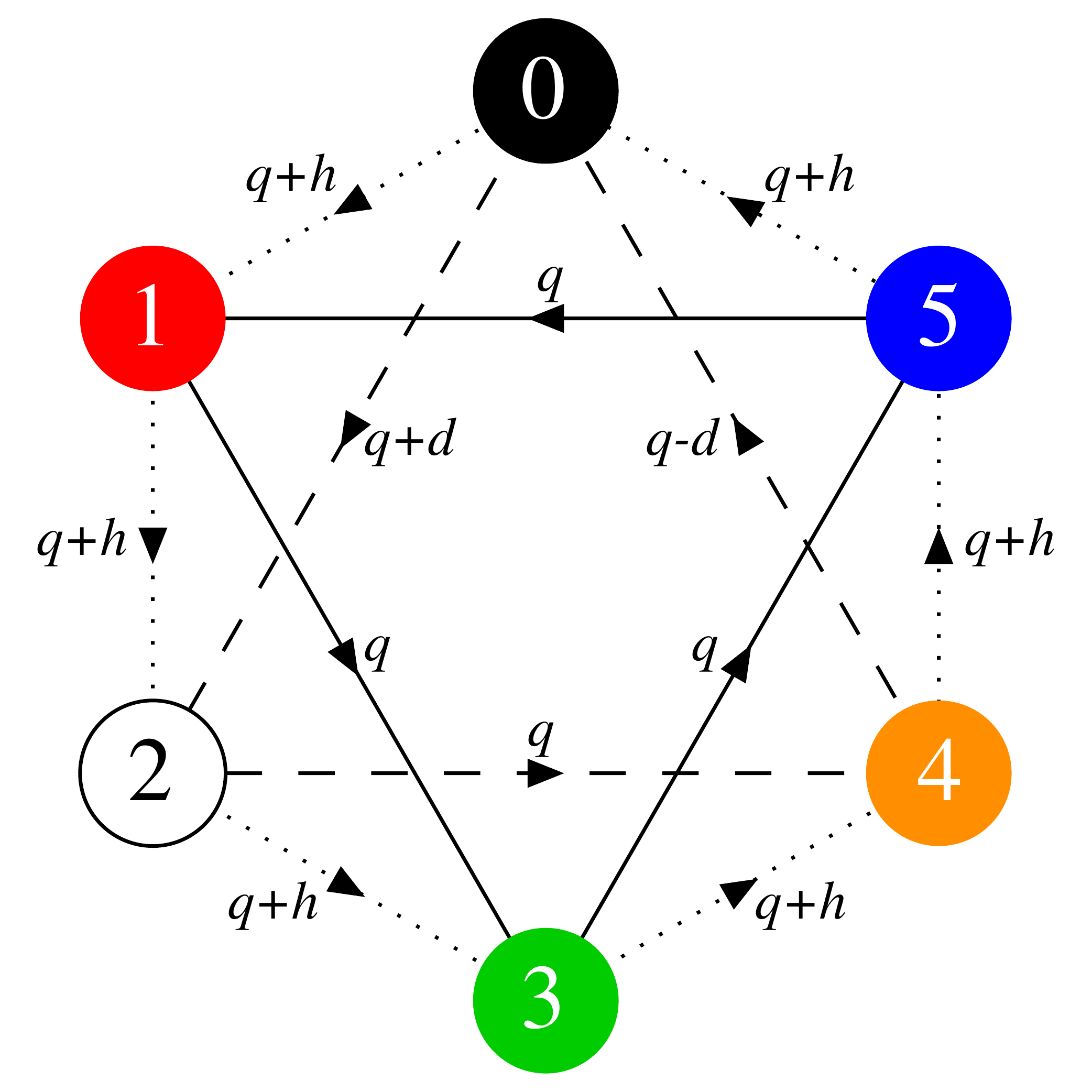}
\caption{\label{model} Food web of the six-member predator-prey model where two three-member defensive alliances compete. While the invasion rates among ${1 \to 3 \to 5}$ species are uniform, the rates among ${0 \to 2 \to 4}$ species are heterogeneous with the same average value.}
\end{center}
\end{figure}

Motivated by these questions, in this work we assume that in one of the competing alliances the invasion rates are different. However, we still prescribe the equal average invasion speed within the heterogeneous loop compared to the value applied in the uniform loop, otherwise the alliance with higher inner invasion rate would have a natural advantage. In our spatial model, that is summarized in Fig.~\ref{model}, six species compete for space on a square grid. Each site $i$ is occupied by one of the six species, hence their distribution is given by a set of site variables $s_i = 0, \dots, 5$. 

\begin{figure*}
\begin{center}
\includegraphics[width=17.0cm]{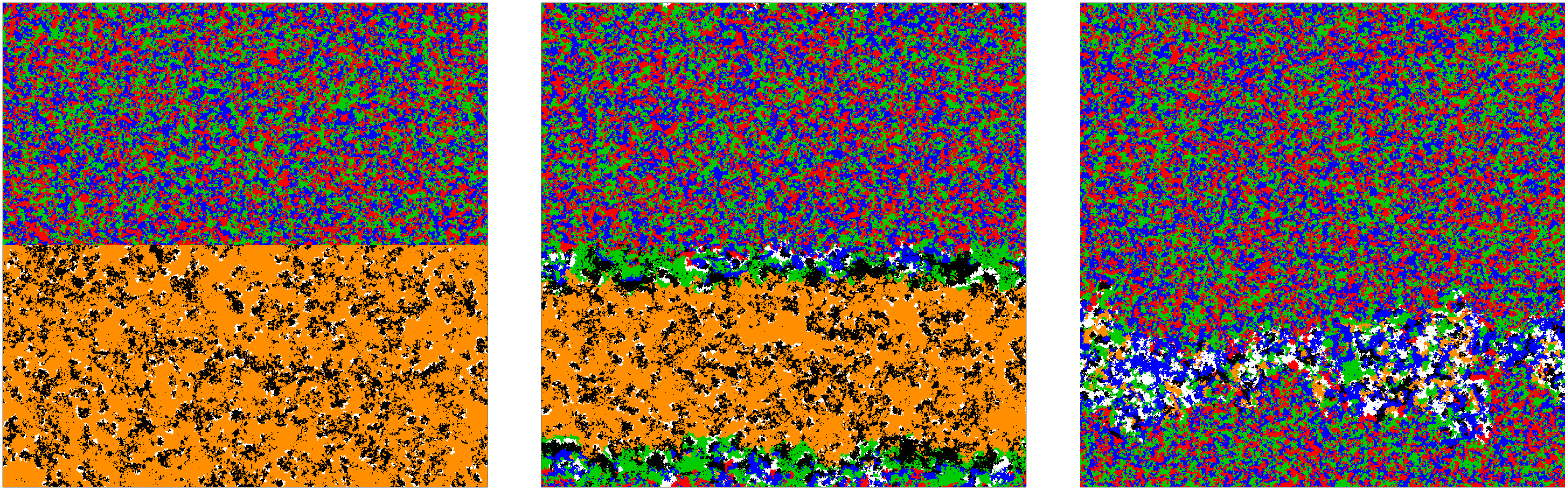}
\caption{\label{battle}Competition of defensive alliances obtained at $q=0.5, h=0.4,$ and $d=0.4$ parameter values. In the beginning we block the strategy invasion across the horizontal borders separating the two halves of the available space. This restriction makes it possible for different solutions (three-species alliances) to emerge independently in the top and bottom regions, respectively. While the domain sizes are almost equal for $(1,3,5)$ alliance on the top, the portions of different color are strikingly unbalanced on the bottom half where the invasion rates are highly heterogeneous among $(0,2,4)$ species. After appropriately long relaxation time we allow solutions to compete by opening the borders for invasions. The middle panel illustrates nicely the vulnerability of the heterogeneous alliance that looses space gradually. In the final stage, not shown, the alliance with uniform invasion rates prevails. The system size is $L \times L = 900 \times 900$ and the snapshots were taken after 100 and 1500 $MC$ steps where 5000 $MC$ steps relaxation period was applied.}
\end{center}
\end{figure*}

According to the food web, odd labeled species form a closed loop  where they all use a uniform $q$ invasion rate among them. Based on the above described mechanism they compose a defensive alliance against even labeled species. The latter $0, 2,$ and $4$ species also form a cyclic dominance, but they apply heterogeneous invasion rates withing their loop. Importantly, the average of their invasion rates is still equal to $q$ value. Here parameter $d>0$ characterizes the diversity of inner invasion rates in the heterogeneous loop. Our third parameter $h$ determines the intensity of external invasion between competing alliances. Evidently, for $h>0$ values the intergroup invasion exceeds the average inner invasion of alliances, while for $h<0$ reduced intergroup invasions happen in the presence of faster average internal invasions.

To reveal the possible consequences of heterogeneous inner invasion rates, we perform Monte Carlo ($MC$) simulations on $L \times L$ square lattice where periodic boundary conditions are used. In an elementary step two nearest neighbors are chosen randomly and if the resident species form a predator-prey pair then an invasion happens and the predator occupies the prey site with the probability determined by the food web plotted in Fig.~\ref{model}. Alternatively, if the chosen neighboring species are neutral, such as species $0$ and $3$, or species are identical then nothing happens. In agreement with the standard simulation protocol, each individual is selected once on average to make a full $MC$ step.

We stress that the frequently used random initial condition, where each species is distributed randomly, can easily result in misleading evolutionary outcome. This is because a random initial condition in a small system does not necessarily provide equal chance for all possible solutions to emerge and compete for space. Starting from a random distribution some solution needs longer relaxation and larger system size to emerge, otherwise alternative solutions block the mentioned solution to be formed. This anomaly is especially valid when the invasion rates are largely different because in this case the typical domain sizes of competing species could be extremely diverse \cite{juul_pre13,vukov_pre13}.

To overcome the mentioned difficulties we apply a specific initial state where different members of defensive alliances are separated in space, which makes it possible for both defensive alliances to emerge independently. During this relaxation period invasions across the borderlines separating alliances are forbidden. After, when both representative patterns have emerged, we open the borders for intergroup invasions and monitor the battle of defensive alliances.

A typical example is given in Fig.~\ref{battle} where we present some representative stages of this competition. Here left panel illustrates the stage when actual fight between defensive alliances starts. Notably, the representative patterns of competing alliances are highly different. This fact, as we will point out later, has a crucial importance on the vitality of an alliance. In particular, on the top half of available space species $1,3,$ and $5$ form domains of equal size. On the bottom half of the lattice, however, the portions of different colors, hence the typical sizes of domains, are largely different, which is a straightforward consequence of heterogeneous invasion rates within the loop \cite{tainaka_pla93}. Interestingly, the largest area of available space is occupied by yellow color, which represents species $4$, who has the smallest invasion rate and predator power against prey species $0$. This is the so-called ``survival of the weakest" effect that characterizes all predator-prey like systems where a cyclic loop is identified \cite{frean_prsb01}. Notably, this feature can also be detected in a broader range of evolutionary game models where dominances between competing strategies are less straightforward, but an effective non-transitive relation may emerge \cite{szolnoki_csf20b}.

The middle panel of Fig.~\ref{battle} illustrates that the alliance formed by red, green, and blue colors gradually invades the region originally occupied by the rival alliance. The right panel depicts the last stage before alliance of $0, 2,$ and $4$ species completely vanishes and the loop formed by $1, 3,$ and $5$ species prevails.

Before analyzing the above described phenomenon, we highlight the importance of appropriately chosen system size and relaxation period which are necessary for the proper stability analysis. More precisely, the half linear size of system should be significantly larger than the typical domain size of each species otherwise we have no chance to produce the requested alliance solution. Naturally, this domain size may depend sensitively on the heterogeneity of invasion rates, therefore we used at least $L \times L = 300 \times 300$ system size, but the majority of simulations were carried out at $L=900$ linear size. However, at specific parameter values we need at least $L=1500$ linear size to obtain the evolutionary outcome that is valid in the large system size limit. Last, we note that similar results can also be obtained when the evolution is launched from a random initial state, but in the latter case we may need a significantly larger system, typically $L=6000$, where all available solutions may emerge somewhere and the most competitive one can gradually invade the whole system. We also stress that the necessary time when one of the solution prevails could be extremely long. For instance even at relatively small system size, say at $L=600$ the requested time to reach the final outcome may exceed $10^7$ $MC$ steps, but already $L=300$ system size may require $5 \cdot 10^5$ steps at specific parameter values.

\begin{figure}
\begin{center}
\includegraphics[width=8.5cm]{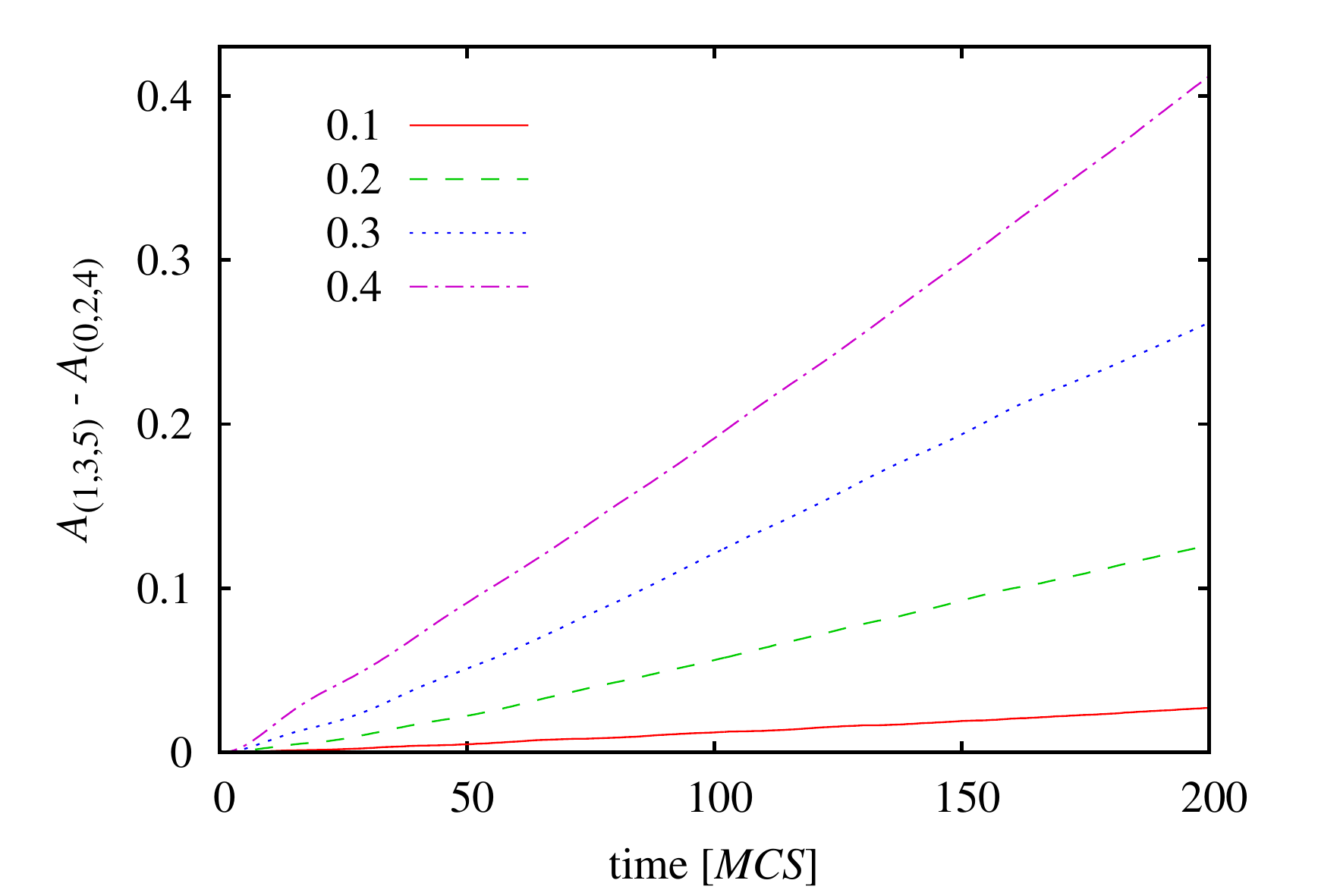}
\caption{\label{homo_d} Evolution of the excess area between the competing alliances as obtained for $q=0.5, h=0.4$ at different 
values of $d$, as indicated in the legend. The linear size of system was $L=900$ where we averaged over 100 independent runs.}
\end{center}
\end{figure}

To characterize the effective intergroup invasion quantitatively we monitor the difference of areas controlled by competing alliances in time. Some representative examples at early stage of evolution are plotted in Fig.~\ref{homo_d} for different values of $d$, as indicated in the legend. These lines suggest that the alliance with homogeneous invasion rates dominate the heterogeneous loop and the strength of invasion, means the slope of curves, depends on the heterogeneity parameter $d$.

\begin{figure}
\begin{center}
\includegraphics[width=8.0cm]{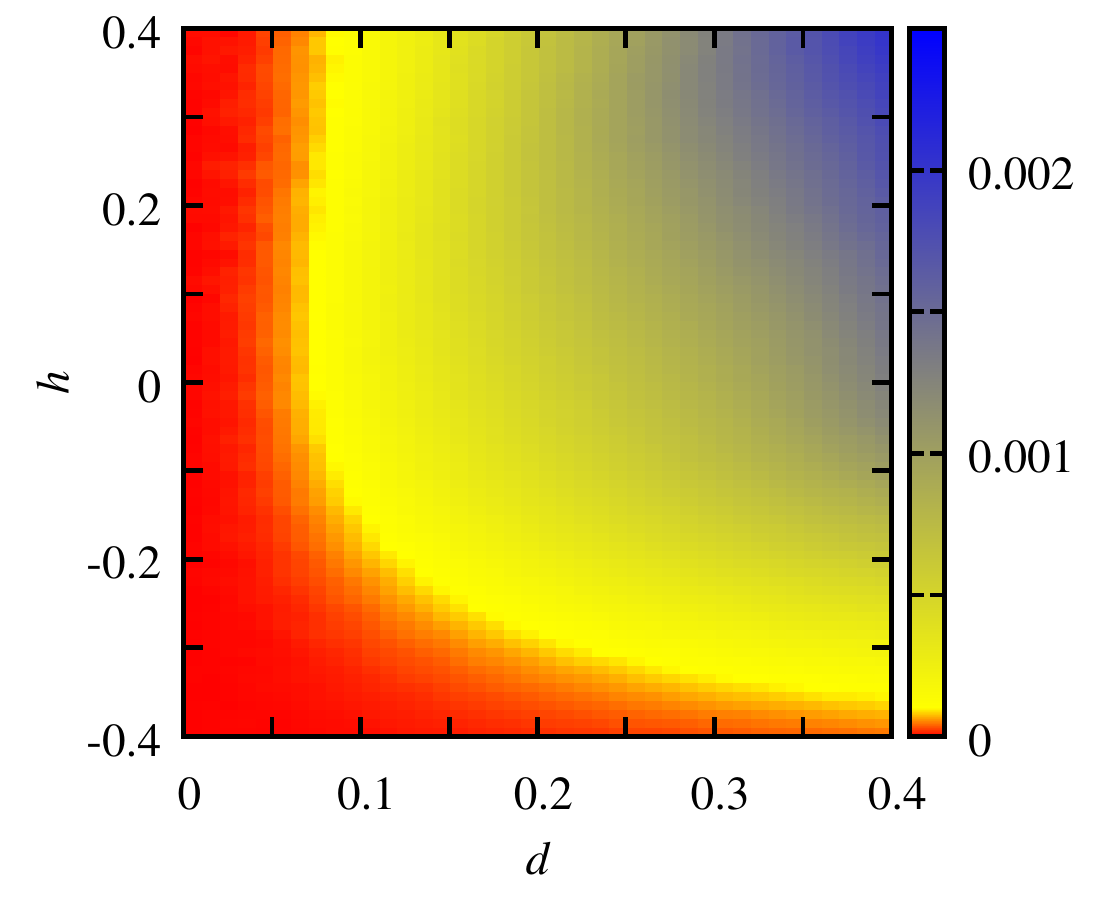}
\caption{\label{scan}The dominance of the alliance using uniform inner invasion rate over the alliance where heterogeneous rates are used at $q=0.5$. The color map encodes the intensity of excess area growth on the $h-d$ parameter plane. The plot suggests that heterogeneous loop becomes  vulnerable especially when the heterogeneity and the external invasion between alliances are high.}
\end{center}
\end{figure}

The general behavior is summarized in Fig.~\ref{scan} where we show the effective invasion rate of homogeneous loop over heterogeneous alliance on the $h-d$ parameter plane. This plot suggests that the strength of dominance of the homogeneous alliance over heterogeneous group increases when the invasion rate between alliances is high (at large $h$) and becomes moderate when this invasion is reduced at large negative $h$ values. However, the effective dominance of homogeneous loop is maintained even at small $h$ values. If we increase the measure of heterogeneity by enlarging $d$ then dominance of homogeneous alliance becomes large, hence the above described vitality of uniform group over heterogeneous alliance is conspicuous at high $(h,d)$ parameter pairs. We note that qualitatively similar behavior can be observed when the average invasion rate is low (at $q=0.2$) or when it is high (for $q=0.8$).

Based on these observations we can conclude that when equal partners form a defensive alliance then it is more vital than a loop where heterogeneous invasion rates are applied in the inner circle. This behavior is related to the emerging pattern due to the presence of heterogeneous invasion rates. Here, the ``weakest" specie, who apply the smallest inner invasion rate, makes the whole alliance vulnerable. This phenomenon is related to the previously mentioned ``survival of the weakest" effect because the ``weakest" species will occupy the majority of available space of the actual alliance. A homogeneous domain, however, is specially sensitive to a rival homogeneous alliance which is characterized by small uniform spots and faster invasion. Hence heterogeneous group members makes the defensive alliance weaker.

One may ask: what if we introduce heterogeneous invasion rates in a non species-specific way? More precisely, we introduce the above used $[q-d,q,q+d]$ invasion rates but these values are randomly used by all species within the $(0,2,4)$ loop. It simply means when a species $0$ invades a neighboring species $2$ then all three values of invasion rate can be used with equal probabilities. Similar dynamical rules are applied to the $2 \to 4$ and to the $4 \to 0$ elementary steps. In this way the average invasion rate remains $q$, but the equality of group members is restored. Interestingly, the vitality of the ``heterogeneous" cycle recovers and this group becomes equal to the $(1,3,5)$ homogeneous association. In other words, when we monitor the excess area of an alliance then the effective slope of this function is zero because both alliances are equally strong now. This feature agrees with the argument we raised about the importance of emerging patterns because when heterogeneous invasion rates are applied in a non species-specific manner then the resulting pattern does not differ relevantly from the domain formation of a homogeneous alliance: since all species are equal in the ``heterogeneous" loop, they occupy equal areas with similar domain sizes as in case of a homogeneous loop.

Until this point we assumed that the average inner invasion rates are equal for both competing alliances. The heterogeneity, more precisely a small invasion rate within a loop, makes the related group vulnerable because unbalanced portions of strategies are easy prey for another alliance where equal partners occupy their territory in a balanced way. It is a straightforward question whether we can support heterogeneous loop by elevating the average inner invasion speed? Previously this quantity was proved to be a decisive factor therefore we may think that an increased inner rotation can cover the weakness caused by the presence of a small invasion rate. To answer this question we extend our model by replacing $d$ parameter with $d_1$ and $d_2$ parameters. More precisely, we assume that in the heterogeneous loop species $0$ invades species $2$ with probability $q+d_1$, while species $4$ invades species $0$ with probability $q-d_2$. In this way by choosing $d_1 > d_2$ parameter pairs we can elevate the average inner speed comparing to the $q$ value applied uniformly in the homogeneous loop.

\begin{figure}
\begin{center}
\includegraphics[width=8.5cm]{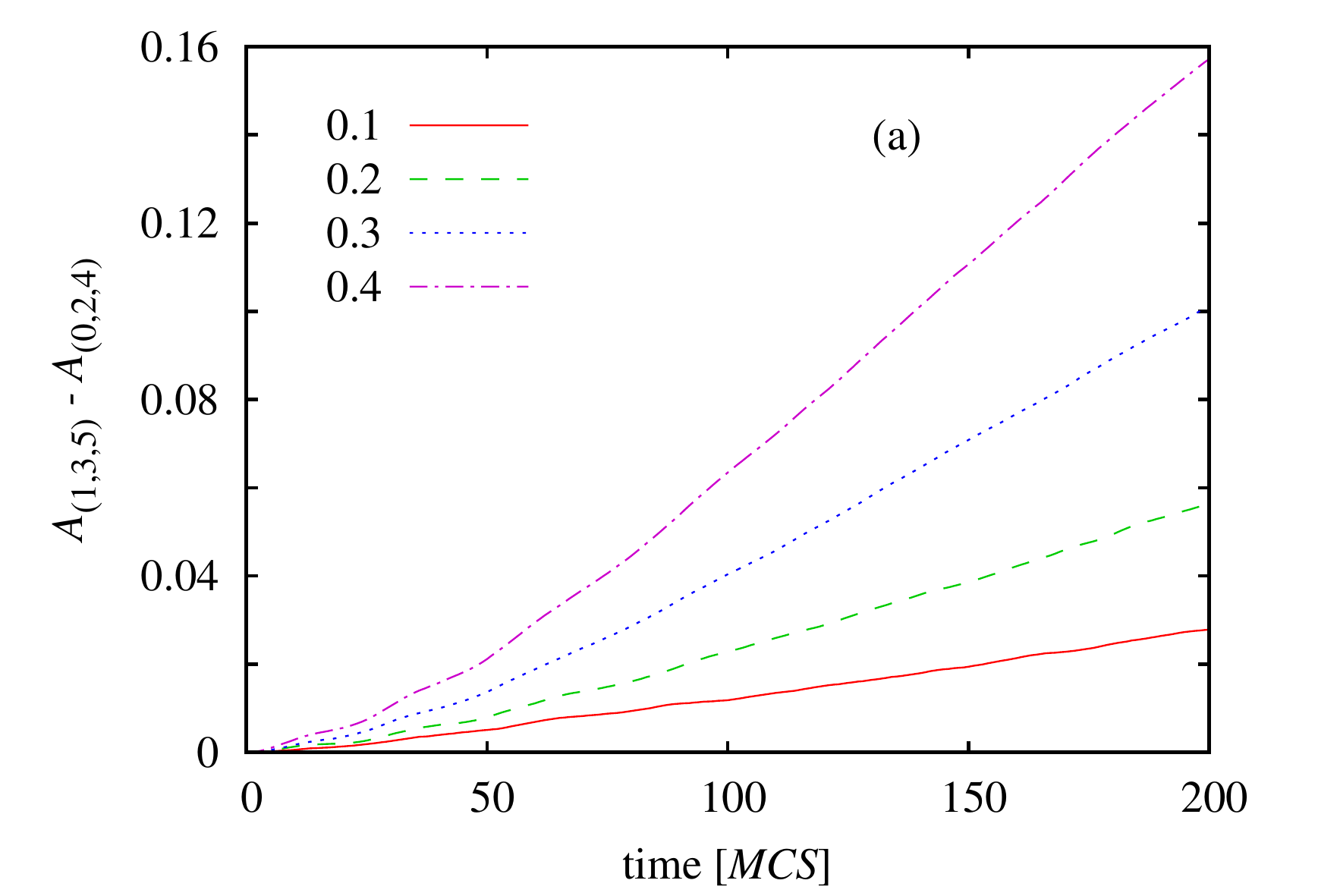}
\includegraphics[width=8.5cm]{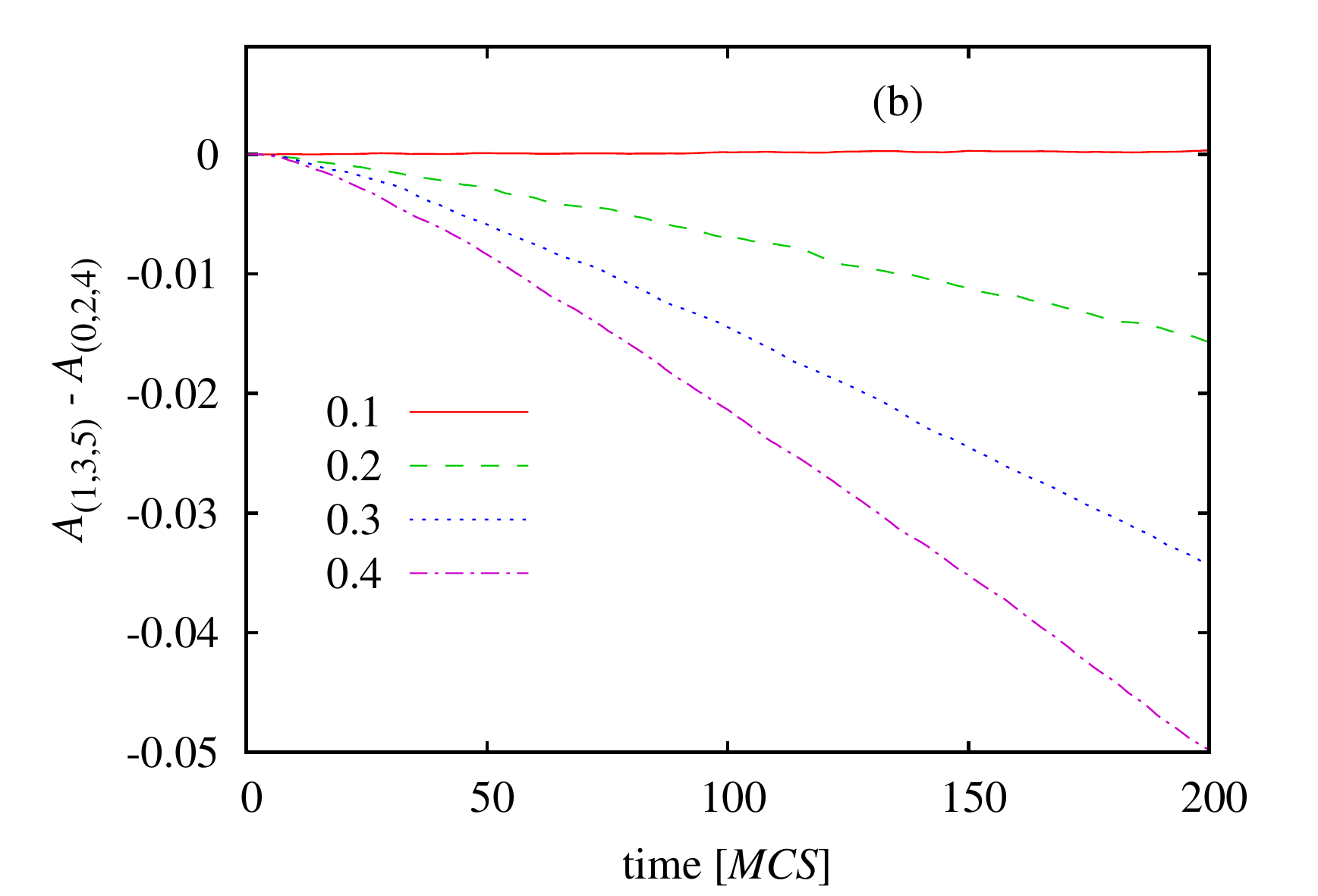}
\caption{\label{d1d2}Competition of alliances in the extended model where heterogeneous loop may have higher average invasion rate. Panel~(a) shows the results when species-specific invasion rates are used at $q=0.2, h=0.4, d_2=0.1$ for different $d_1$ values, as indicated in the legend. Panel~(b) depicts the results when an invasion rate is not dedicated to a species, but the whole available set can be applied by all members of the alliance. Here, when equality is restored among group members, the higher average invasion rate provides benefit to the loop. Parameter values are $q=0.5, h=0.4, d_2=0.1$, while different $d_1$ values are indicated in the legend.}
\end{center}
\end{figure}

To our great surprise, the dominance of homogeneous loop remains intact for the majority of parameter space no matter the applied $(d_1,d_2)$ parameter pair can ensure a significantly faster inner invasion. Some examples are given in panel~(a) of Fig.~\ref{d1d2} where we plotted the evolution of excess area for different $d_1$ values at a fixed $d_2$ value. Evidently, larger $d_1$ means faster average rotation within the heterogeneous loop. Nevertheless, the system reaction is the opposite of the behavior one may naively expect. As the mentioned panel illustrates, the dominance of homogeneous loop can be more pronounced for larger $d_1$ values when inner invasion is faster in the heterogeneous alliance. Put differently, the latter loop becomes even more sensitive when we increase the difference between $d_1$ and $d_2$. Based on the argument we gave about the importance of balanced domain distribution, this behavior can be understood, because larger $d_1$ results in more heterogeneous specie-specific invasion rates, hence large uniform domains of a particular species. And the latter makes the alliance weaker.

Previously we pointed out that the usage of heterogeneous invasion rates in a non species-specific way does not hurt the equality of species therefore such modification is harmless. But if it is the case, then speeding up the inner rotation should have a positive consequence on the vitality of an alliance. This argument is nicely confirmed by panel~(b) of Fig.~\ref{d1d2} where we used different $d_1$ values at a fixed $d_2$ value. First, when $d_1=d_2$ (top curve) then there is no increment of excess area of any alliances. As we already reported, in this case the alliance which uses uniform invasion rates and the alliance where heterogeneous rates are applied remain equally strong. Secondly, for $d_1>d_2$ values the curves have negative slope indicating the dominance of even labeled species. This means that speeding up the inner rotation does have the expected consequence on the vitality of the loop using heterogeneous rates. Last, the curves become steeper as we increase $d_1$ which is again in nice agreement with the general expectation about the role of inner invasion speed.

Turning back to the species-specific version of the extended model, the victory of the homogeneous loop is unambiguous for fast inner rotations when the value of $q$ is high. For moderate $q$ values, however, the unexpected dominance of homogeneous loop cannot be maintained for the whole parameter space. In the latter case there are some limited regions where the alliance using heterogeneous rates can prevail or other type of solutions dominate the homogeneous loop. 

\begin{figure}[h!]
\begin{center}
\includegraphics[width=8.5cm]{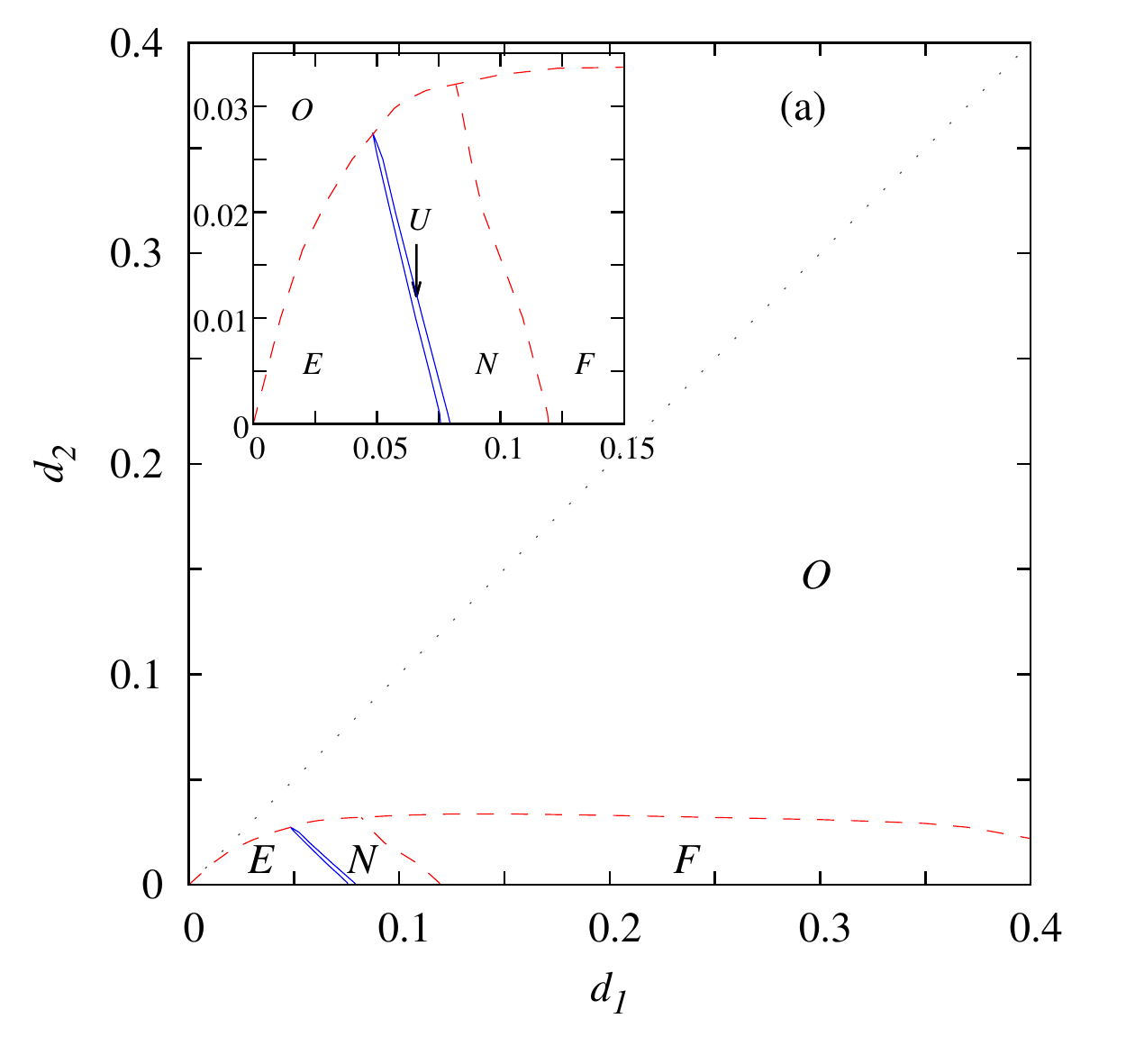}
\includegraphics[width=8.5cm]{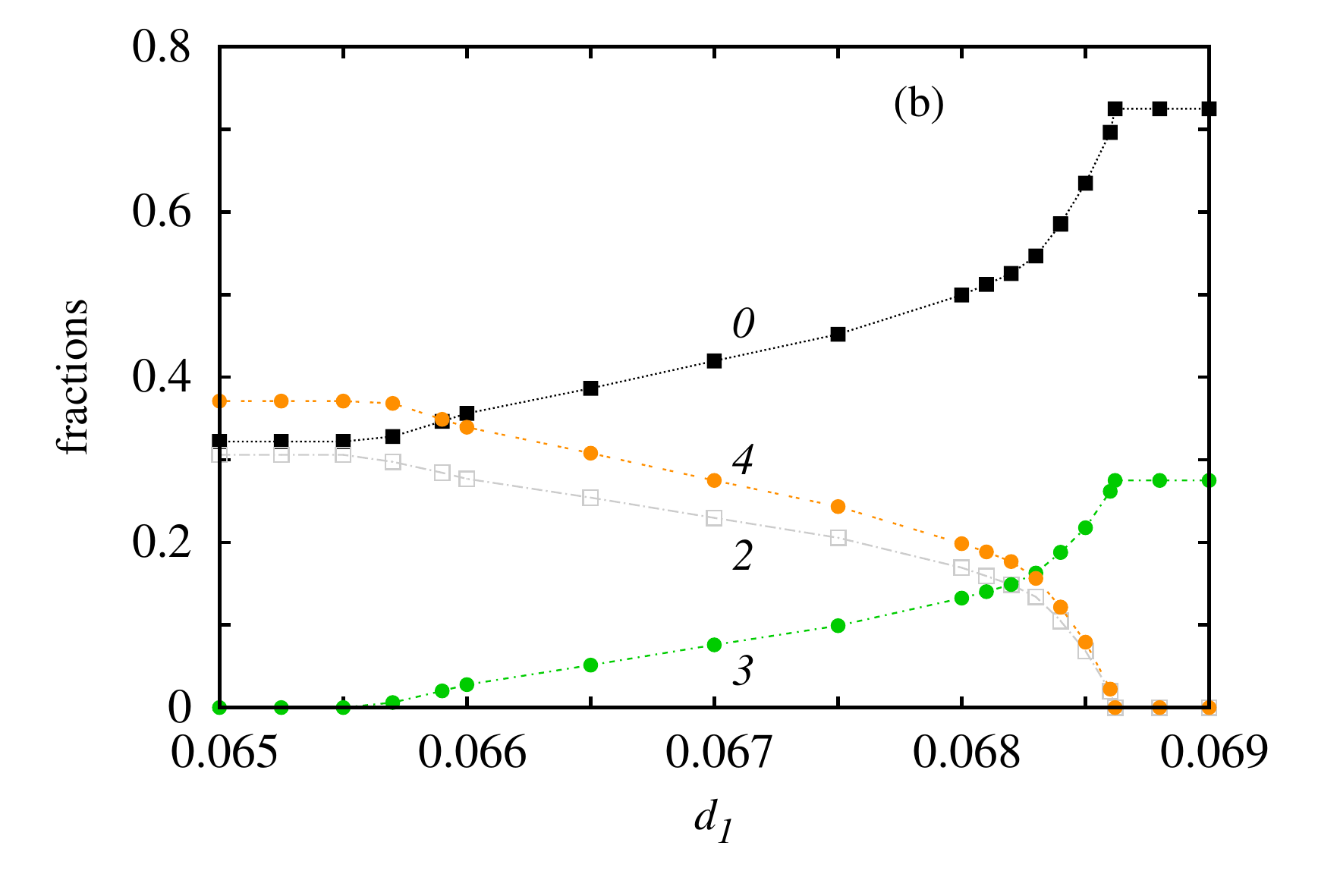}
\caption{\label{phd}Panel~(a) depicts a representative phase diagram of the extended model in the moderate average invasion rate region obtained at $q=0.4, h=0.4$. The diagonal $d_1=d_2$ dotted line marks the border where the average inner invasion speeds are equal for the homogeneous and heterogeneous loops. The inset shows the enlarged part of the diagram for small $d_2$ values. Here ``$O$" denotes the phase where homogeneous alliance composed by $(1+3+5)$ species dominates. ``$E$" marks the parameter region where the heterogeneous loop of $(0+2+4)$ species prevails. Label ``$F$" depicts the region where all species except species $1$ survive, while label $N$ denotes the area where only neutral ($0+3$) pairs remain alive during the evolution. Interestingly, there is a tiny region, marked by ``$U$", where a four-species solution emerges. This is the heterogeneous alliance extended by species $3$.
The latter phase is illustrated in panel~(b) where the stationary fractions of surviving species are shown in dependence of the control parameter $d_1$ at a fixed $d_2 = 0.01$ value.}
\end{center}
\end{figure}

A representative plot of phase diagrams is shown in Fig.~\ref{phd}~(a) where we present the stable solutions on the $d_1 - d_2$ parameter plane at fixed $q=0.4, h=0.4$ values. We note that the diagonal $d_1 = d_2$ dotted line marks the border where the average inner invasions are equal for both alliances. To the right from this line the $d_1 > d_2$ condition ensures a faster inner rotation for the heterogeneous loop. Still, the dominance of the homogeneous group remains intact for almost everywhere. This phase, where species $1+3+5$ dominate, is marked by ``$O$" in the diagram. As the diagram depicts, the system terminates onto alternative solutions only for small $d_2$ values. This parameter space is zoomed in the inset of panel~(a) of Fig.~\ref{phd}. 

At small $d_2$, when there is no reasonable difference between the smallest invasion rates in the heterogeneous loop, we can distinguish four additional phases in dependence of the control parameter. When $d_1$ is small, the heterogeneous alliance, composed by species $(0+2+4$), is capable to dominate the homogeneous loop. This phase is marked by ``$E$". As we previously mentioned, this behavior would be the expected evolutionary outcome based on the enhanced rotation of the inner cycle, but this is justified at a very limited parameter space only.
 
Instead, by increasing $d_1$ the system evolves into a state where only species $0$ and its neutral partner, species $3$ survive. This phase is denoted by ``$N$" (neutral) on the phase diagram. Interestingly, between the mentioned phases there is a very narrow parameter space where a new kind of solution emerges. In the latter phase, which is marked by ``$U$", four species coexist. These are the member of the heterogeneous $(0+2+4)$ alliance extended by species $3$, who is the mentioned neutral pair of species $0$. Panel~(b) of Fig.~\ref{phd}, where we plotted the stationary factions of surviving species, illustrates clearly that there are continuous phase transitions towards phase ``$E$" and to phase ``$N$" when we decrease or increase the value of control parameter $d_1$. We note that to measure properly the stationary states in this intermediate ``$U$" phase required at least $L=1500$ linear system size due to the large fluctuations in the vicinity of transition points. 

As we increase $d_1$ even further, there is a discontinuous phase transition between phase ``$N$" and phase ``$F$". The latter denotes a new solution where five species of the whole six competitors coexist permanently. The only missing rival is species $1$ who is the external invader of the elementary process which happens with the highest probability. Indeed, because of the relatively high value of $d_1$, the $0 \to 2$ inner invasion happens so fast in the domain of heterogeneous alliance that paralyzes species $1$. As a result, the latter has no proper chance to invade species $2$ because the inner invasion by species $0$ happens fast, hence species $1$ meets its predator $0$ more frequently. Notably, similar effect cannot hinder species $3$ and $5$ because of  the moderate value of $q$ and $q-d_2$. This explains why two external species can coexist with the highly heterogeneous $(0,2,4)$ alliance.

It is not visible in the phase diagram plotted in Fig.~\ref{phd}~(a), but it is a general behavior that by increasing $d_1$ further the system reaches phase ``$O$" again. This phenomenon can be explained by the argument we already mentioned. Here the difference within the heterogeneous alliance becomes so large, which results in highly different sizes of domains of group members. The latter makes a group always vulnerable against a loop where equal partners maintain a homogeneous domain pattern.

Summing up, in this Letter we emphasized the paramount importance of balanced invasion rates in a defensive alliance. When group members are not equally strong, or there is significant difference between them, then such alliance cannot function efficiently. In the latter case such formation may be dominated by a well-balanced alliance or external invaders can coexist with the highly heterogeneous defensive alliance. In other words, when the differences within a closed loop are too high then the concept of defensive alliance does not work properly. This effect is robust and remains valid if we change the interaction topology and replace von Neumann neighborhood by Moore neighborhood where every individual competes directly with 8 neighbors.

We stress that our observations are not restricted to Lotka-Volterra type population dynamical models \cite{avelino_epl18,nagatani_pa19b,park_c18c,bazeia_pre19,nagatani_jtb19,takeshue_epl19,park_c19b}, but can be applied to evolutionary game models, too. In the latter case non-transitive interaction between elementary strategies may emerge easily especially when several strategies compete and the concept of alliances could be a decisive to understand the evolutionary outcomes \cite{szolnoki_pre17,canova_jsp18,mobilia_g16,szolnoki_pre10b,li_xp_pla20,brown_pre19}.

In present work we only focused on species-specific and non-species specific usage of heterogeneous invasion rates, but there are other options which challenge future research efforts. Just to give an example, spatial models of structured populations make realistic the usage of site-specific invasion rates in a heterogeneous environment \cite{szolnoki_srep16, roman_pre13, shao_yx_epl19, wang_x_rspa20} or the presence of zealots may also influence the elementary invasions \cite{mobilia_prl03, szolnoki_pre16, mobilia_jsm07, szolnoki_njp15,cardillo_prr20}. Further related research avenue could be the mobility of players \cite{reichenbach_n07} or when we go beyond pairwise interaction because the presence of a third party could modify the elementary invasion process  \cite{kelsic_n15, szolnoki_pre14c, battiston_pr20}.

\begin{acknowledgments}
This research was supported by the Hungarian National Research Fund (Grant K-120785).
\end{acknowledgments}


\begin{thebibliography}{10}
\expandafter\ifx\csname url\endcsname\relax\def\url#1{\texttt{#1}}\fi

\bibitem{kerr_n02}
\Name{Kerr B., Riley M.~A., Feldman M.~W. \and Bohannan B. J.~M.}
  \REVIEW{Nature}{418}{2002}{171}.

\bibitem{szolnoki_jrsif14}
\Name{Szolnoki A., Mobilia M., Jiang L.-L., Szczesny B., Rucklidge A.~M. \and
  Perc M.} \REVIEW{J. R. Soc. Interface}{11}{2014}{20140735}.

\bibitem{baker_jtb20}
\Name{Baker R. \and Pleimling M.} \REVIEW{J. Theor. Biol.}{486}{2020}{110084}.

\bibitem{frey_pa10}
\Name{Frey E.} \REVIEW{Physica A}{389}{2010}{4265}.

\bibitem{mathiesen_prl11}
\Name{Mathiesen J., Mitarai N., Sneppen K. \and Trusina A.} \REVIEW{Phys. Rev.
  Lett.}{107}{2011}{188101}.

\bibitem{nagatani_srep18}
\Name{Nagatani T., Ichinose G. \and i.~Tainaka K.} \REVIEW{Sci. Rep.}{8}{2018}{7094}.

\bibitem{dobramysl_jpa18}
\Name{Dobramysl U., Mobilia M., Pleimling M. \and T{\"a}uber U.~C.} \REVIEW{J.
  Phys. A: Math. Theor.}{51}{2018}{063001}.

\bibitem{park_c18}
\Name{Park J.} \REVIEW{Chaos}{28}{2018}{053111}.

\bibitem{avelino_pre19b}
\Name{Avelino P.~P., \protect{de Oliveira} B.~F. \and Trintin R.~S.}
  \REVIEW{Phys. Rev. E}{100}{2019}{042209}.

\bibitem{watt_je47}
\Name{Watt A.~S.} \REVIEW{J. Ecol.}{35}{1947}{1}.

\bibitem{lankau_s07}
\Name{Lankau R.~A. \and Strauss S.~Y.} \REVIEW{Science}{317}{2007}{1561}.

\bibitem{jackson_pnas75}
\Name{Jackson J. B.~C. \and Buss L.} \REVIEW{Proc. Nat. Acad. Sci. USA}{72}{1975}{5160}.

\bibitem{sinervo_n96}
\Name{Sinervo B. \and Lively C.~M.} \REVIEW{Nature}{380}{1996}{240}.

\bibitem{hauert_s02}
\Name{Hauert C., De~Monte S., Hofbauer J. \and Sigmund K.} \REVIEW{Science}{296}{2002}{1129}.

\bibitem{semmann_n03}
\Name{Semmann D., Krambeck H.-J. \and Milinski M.} \REVIEW{Nature}{425}{2003}{390}.

\bibitem{helbing_ploscb10}
\Name{Helbing D., Szolnoki A., Perc M. \and Szab{\'o} G.} \REVIEW{PLoS Comput.
  Biol.}{6}{2010}{e1000758}.

\bibitem{sigmund_pnas01}
\Name{Sigmund K., Hauert C. \and Nowak M.~A.} \REVIEW{Proc. Natl. Acad. Sci.
  USA}{98}{2001}{10757}.

\bibitem{szolnoki_epl10}
\Name{Szolnoki A. \and Perc M.} \REVIEW{EPL}{92}{2010}{38003}.

\bibitem{requejo_pre12}
\Name{Requejo R.~J. \and Camacho J.} \REVIEW{Phys. Rev. E}{85}{2012}{066112}.

\bibitem{szabo_pr07}
\Name{Szab{\'o} G. \and F{\'a}th G.} \REVIEW{Phys. Rep.}{446}{2007}{97}.

\bibitem{roman_jtb16}
\Name{Roman A., Dasgupta D. \and Pleimling M.} \REVIEW{J. Theor. Biol.}{403}{2016}{10}.

\bibitem{szabo_pre01b}
\Name{Szab{\'o} G. \and Cz{\'a}r{\'a}n T.} \REVIEW{Phys. Rev. E}{64}{2001}{042902}.

\bibitem{perc_pre07b}
\Name{Perc M., Szolnoki A. \and Szab{\'o} G.} \REVIEW{Phys. Rev. E}{75}{2007}{052102}.

\bibitem{szolnoki_epl15}
\Name{Szolnoki A. \and Perc M.} \REVIEW{EPL}{110}{2015}{38003}.

\bibitem{tainaka_epl91}
\Name{Tainaka K. \and Itoh Y.} \REVIEW{Europhys. Lett.}{15}{1991}{399}.

\bibitem{juul_pre13}
\Name{Juul J., Sneppen K. \and Mathiesen J.} \REVIEW{Phys. Rev. E}{87}{2013}{042702}.

\bibitem{vukov_pre13}
\Name{Vukov J., Szolnoki A. \and Szab{\'o} G.} \REVIEW{Phys. Rev. E}{88}{2013}{022123}.

\bibitem{tainaka_pla93}
\Name{Tainaka K.} \REVIEW{Phys. Lett. A}{176}{1993}{303}.

\bibitem{frean_prsb01}
\Name{Frean M. \and Abraham E.~D.} \REVIEW{Proc. R. Soc. Lond. B}{268}{2001}{1323}.

\bibitem{szolnoki_csf20b}
\Name{Szolnoki A. \and Chen X.} \REVIEW{Chaos Soliton. Fract.}{138}{2020}{109935}.

\bibitem{avelino_epl18}
\Name{Avelino P.~P., Bazeia D., Losano L., Menezes J. \and \protect{de
  Oliveira} B.~F.} \REVIEW{EPL}{121}{2018}{48003}.

\bibitem{nagatani_pa19b}
\Name{Nagatani T.} \REVIEW{Physica A}{525}{2019}{1114}.

\bibitem{park_c18c}
\Name{Park J., Do Y. \and Jang B.} \REVIEW{Chaos}{28}{2018}{113110}.

\bibitem{bazeia_pre19}
\Name{Bazeia D., \protect{de Oliveira} B.~F. \and Szolnoki A.} \REVIEW{Phys.
  Rev. E}{99}{2019}{052408}.

\bibitem{nagatani_jtb19}
\Name{Nagatani T., Ichinose G. \and i.~Tainaka K.} \REVIEW{J. Theor. Biol.}{462}{2019}{425}.

\bibitem{takeshue_epl19}
\Name{Takeshue H.} \REVIEW{EPL}{126}{2019}{58001}.

\bibitem{park_c19b}
\Name{Park J. \and Jang B.} \REVIEW{Chaos}{29}{2019}{051105}.

\bibitem{szolnoki_pre17}
\Name{Szolnoki A. \and Chen X.} \REVIEW{Phys. Rev. E}{95}{2017}{052316}.

\bibitem{canova_jsp18}
\Name{Canova G.~A. \and Arenzon J.~J.} \REVIEW{J. Stat. Phys.}{172}{2018}{279}.

\bibitem{mobilia_g16}
\Name{Mobilia M., Rucklidge A.~M. \and Szczesny B.} \REVIEW{Games}{7}{2016}{24}.

\bibitem{szolnoki_pre10b}
\Name{Szolnoki A., Wang Z., Wang J. \and Zhu X.} \REVIEW{Phys. Rev. E}{82}{2010}{036110}.

\bibitem{li_xp_pla20}
\Name{Li X., Wang H., Hao G. \and Xia C.} \REVIEW{Phys. Lett. A}{384}{2020}{125414}.

\bibitem{brown_pre19}
\Name{Brown B.~L., Meyer-Ortmanns H. \and Pleimling M.} \REVIEW{Phys. Rev. E}{99}{2019}{062116}.

\bibitem{szolnoki_srep16}
\Name{Szolnoki A. \and Perc M.} \REVIEW{Sci. Rep.}{6}{2016}{23633}.

\bibitem{roman_pre13}
\Name{Roman A., Dasgupta D. \and Pleimling M.} \REVIEW{Phys. Rev. E}{87}{2013}{032148}.

\bibitem{shao_yx_epl19}
\Name{Shao Y., Wang X. \and Fu F.} \REVIEW{EPL}{126}{2019}{40005}.

\bibitem{wang_x_rspa20}
\Name{Wang X., Zheng Z. \and Fu F.} \REVIEW{Proc. R. Soc. A}{476}{2020}{20190643}.

\bibitem{mobilia_prl03}
\Name{Mobilia M.} \REVIEW{Phys. Rev. Lett.}{91}{2003}{028701}.

\bibitem{szolnoki_pre16}
\Name{Szolnoki A. \and Perc M.} \REVIEW{Phys. Rev. E}{93}{2016}{062307}.

\bibitem{mobilia_jsm07}
\Name{Mobilia M., Petersen A. \and Redner S.} \REVIEW{J. Stat. Mech.}{2007}{2007}{P08029}.

\bibitem{szolnoki_njp15}
\Name{Szolnoki A. \and Perc M.} \REVIEW{New J. Phys.}{17}{2015}{113033}.

\bibitem{cardillo_prr20}
\Name{Cardillo A. \and Masuda N.} \REVIEW{Phys. Rev. Res.}{2}{2020}{023305}.

\bibitem{reichenbach_n07}
\Name{Reichenbach T., Mobilia M. \and Frey E.} \REVIEW{Nature}{448}{2007}{1046}.

\bibitem{kelsic_n15}
\Name{Kelsic E.~D., Zhao J., Vetsigian K. \and Kishony R.} \REVIEW{Nature}{521}{2015}{516}.

\bibitem{szolnoki_pre14c}
\Name{Szolnoki A., Vukov J. \and Perc M.} \REVIEW{Phys. Rev. E}{89}{2014}{062125}.

\bibitem{battiston_pr20}
\Name{Battiston F., Cencetti G., Iacopini I., Latora V., Lucas M., Patania A.,
  Young J.-G. \and Petri G.} \REVIEW{Phys. Rep.}{}{2020}{doi:10.1016/j.physrep.2020.05.004}.

\end{thebibliography}
\end{document}